\documentclass[useAMS,a4paper,fleqn,usenatbib]{mnras}

\usepackage{newtxtext,newtxmath}

\usepackage[T1]{fontenc}
\usepackage{ae,aecompl}

\usepackage{graphicx}	
\usepackage{amsmath}	
\usepackage{amssymb}	
\usepackage{url}
\usepackage{color}
\usepackage{tabularx}

\mathchardef\mhyphen="2D

\newcommand{\civ}{\ifmmode {\rm C}\,{\sc iv} \else C\,{\sc iv}\fi}
\newcommand{\CIV}{\ifmmode {\rm C}\,{\sc iv}\,\lambda1549 \else 
	           C\,{\sc iv}\,$\lambda1549$\fi}
\newcommand{\oiii}{O\,{\sc iii}}

\newcommand{\mgii}{Mg\,{\sc ii}}

\newcolumntype{L}[1]{>{\raggedright\arraybackslash}p{#1}}
\newcolumntype{C}[1]{>{\centering\arraybackslash}p{#1}}
\newcolumntype{R}[1]{>{\raggedleft\arraybackslash}p{#1}}

\title[Orientation effects on Torus emission]{Orientation effects on the near-infrared broad band emission of quasars}

\author[S. Bisogni et al.]{
Susanna Bisogni,$^{1,2,3}$\thanks{E-mail: susanna@arcetri.astro.it}
Elisabeta Lusso,$^{4}$
Alessandro Marconi,$^{1,3}$
and Guido Risaliti$^{1,3}$
\\
$^{1}$Dipartimento di Fisica e Astronomia, Universit\`a degli Studi di Firenze, Via. G. Sansone 1, 50019 Sesto Fiorentino (FI), Italy,\\
$^{2}$Harvard-Smithsonian Center for Astrophysics, 60 Garden Street, Cambridge, MA 02138, USA\\
$^{3}$INAF - Osservatorio Astrofisico di Arcetri, Largo Enrico Fermi 5, I-50125 Firenze, Italy \\
$^{4}$Centre for Extragalactic Astronomy, Durham University, South Road, Durham, DH1 3LE, UK
}

\date{Accepted XXX. Received YYY; in original form ZZZ}

\pubyear{2019}

\begin{document}
\label{firstpage}
\pagerange{\pageref{firstpage}--\pageref{lastpage}}
\maketitle

\begin{abstract}
We recently proposed the equivalent width (EW) of the narrow [OIII]5007\AA~ emission line as an orientation indicator for active galactic nuclei. We tested this method on about 12,300 optically selected broad line quasars from the SDSS 7th Data Release at redshift $z<0.8$ and with full width at half-maximum values of broad emission lines (H$\alpha$, H$\beta$, and \mgii) larger than 2000 km/s. We now examine their spectral energy distributions (SEDs) using broad band photometry from the near-infrared to the ultraviolet to look for variations in the overall shape as a function of the EW[OIII]. We find that quasars with low EW[OIII] values (close to face-on position) have flatter near-infrared SEDs with respect to sources with high EW[OIII] values (almost edge-on). Moreover, quasars with high EW[OIII] values show a factor of $\sim$2 lower emission in the UV to quasars with low EW[OIII] values. Our findings indicate that the torus is clumpy and, on average, co-axial with the accretion disc and broad line region, in agreement with the most recent theoretical models for the obscuring torus.
\end{abstract}




\section{Introduction}
\label{sec:introduction}

We proposed the equivalent width (EW) of narrow [\oiii] emission line as an orientation indicator for quasars \citep[][hereafter Paper I]{Risaliti2011, Bisogni2017}. The anisotropy of the emission coming from the accretion disc \citep{ShakuraSunyaev1973} and the properties of the [\oiii] line at $5007$\AA~ emitted by the Narrow Line Region (NLR) (isotropic and ascribable for the most part to AGN activity \citep{Mulchaey1994, Heckman2004} allow us to use EW[\oiii] as an indicator of source inclination with respect to the line-of-sight.

To test our method, we selected a large sample ($\sim 12,300$ objects) of unobscured type 1 quasars from the SDSS DR7 \citep{Schneider2010,Abazajian2009}, divided it in bins of EW[\oiii] and created, for each one of them, a stacked spectrum meant to be representative of the bin (Paper~I).
All quasars have broad full width at half maximum values larger than 2000 km/s of the most prominent broad emission lines (i.e. H$\alpha$, H$\beta$, and \mgii).
The presence of orientation effects in both the broad and the narrow emission lines of optical/UV stacked spectra as a function of EW[\oiii] has then proven the reliability of this quantity as an orientation indicator.
Specifically, in the case of the lines emitted by the Broad Line Region (BLR, e.g. H$\alpha$, H$\beta$, and \mgii) we recognised a broadening of the line profiles when moving from low EW[\oiii] objects, corresponding to mostly face-on sources, to high EW[\oiii] sources, corresponding to edge-on sources, in agreement with BLR models that describe the geometry of this structure as disc-shaped \citep{Pancoast2014, Grier2017}. 
In the case of lines emitted by the NLR, represented by the [\oiii] $5007$\AA~ line itself, we observed a statistically significant decrease both in the shift of the central velocity with respect to the reference wavelength and in the intensity of the blue component of the line when moving from low to high EW[\oiii], in agreement with the presence of gas in outflow in this region, that we are able to intercept preferentially when the source is face-on.
The presence of such effects on spectroscopic features confirms that the orientation of the source with respect to the line of sight is one of the main drivers of the variance of quasars spectra, although not the only one \citep[SMBH mass, Eddington ratio, etc. , e.g. ][]{Marziani2003, ShenHo2014}.\\

\vspace{0.01cm}
We now investigate the quasar broad band spectral energy distributions (SEDs) from the near-infrared to the ultraviolet to look for variations in the overall shape as a function of the EW[OIII]. Specifically, we focus on the near-infrared emission, which is usually ascribed to a dusty structure surrounding the disc at parsec scales (the so-called torus) proposed in the orientation empirical model \citep{Antonucci1993, UrryPadovani1995}.
This component, coplanar with the accretion disc but with much larger scale height \citep{Krolik1988, Hicks2009} to hide the inner components, has been recently proposed to be characterized by a clumpy \citep[e.g. ][]{Nenkova2008a, Nenkova2008b, Nikutta2009, HoenigKishimoto2010, Stalevsky2012, Siebenmorgen2015}, rather than smooth \citep{PierKrolik1992, Fritz2006} structure.
The clumpy description fits well with the results of Paper~I: given the selection we applied to our sample, only blue, type 1 objects are included; the fact that we can intercept the accretion disc and BLR emission even in edge-on positions is therefore a confirmation that - if present - the obscuring component can not be at the same time smooth and coplanar with the inner components. Only three scenarios are then plausible: 1. an obscuring structure not present at all, 2. the torus is misaligned with respect to the accretion disc and the BLR or 3. the torus is clumpy and we can have a line of sight of the inner regions even if they are coplanar.
An analysis of the emission in the IR band of the same sample as a function of EW[\oiii] is therefore required in order to have information on the properties of the torus. We analyse the SED of the sources in the sample using the UV/optical/IR photometry available in literature, then focusing on the part of the SED ascribable to the torus.

This paper is organized as follows: in Section \ref{sec:data} we describe the data, in Section \ref{sec:data_analysis} we discuss the analysis we performed interpolating the photometric data available. In Sections \ref{sec:results} and \ref{sec:discussion} our results and their implications on the properties of the torus are presented and discussed.
In this work, the following cosmological parameters are assumed: $H_{0} = 70$ km s$^{-1}$ Mpc$^{-1}$, $\Omega_{M}=0.3$ and $\Omega_{\Lambda}=0.7$.

\section{Data} \label{sec:data}

\subsection{The sample} \label{subsec:sample}
Our sample was selected from the Quasar Catalogue of the Sloan Digital Sky Survey (SDSS) DR7 \citep{Schneider2010} - through the properties compiled in the catalogue by \cite{Shen2011} - to be firstly composed by objects whose [\oiii] emission falls in the optimal response window of the SDSS spectrograph ($0.001 < z < 0.8$). We then selected quasars with an absolute magnitude $M_{i}<-22.1$ and a median signal-to-noise per pixel for the rest-frame $4750 \mhyphen 4950$\AA~region S/N$\geq 5$. The final requirement was for the EW[\oiii] to be in the range $1 \mhyphen 300$\AA~\citep{Risaliti2011}, so to exclude sources with an incorrect measure of the [\oiii] line. This selection yielded a sample of $\sim 12,300$ quasars.

\subsection{Photometric data}
We collected broad band photometry from IR to the UV using data from all sky surveys, such as \emph{WISE}, 2MASS, SDSS and \emph{GALEX}, to allow a proper determination of the emission coming from the accretion disk and the obscuring torus.

The \emph{Galaxy Evolution Explorer} (\emph{GALEX}) is an ultraviolet space telescope that performed a survey in the FUV ($1350-1780$\AA) and NUV ($1770-2730$\AA) ultraviolet bands.
We used the \emph{GALEX} photometric data from the sixth release (GR5) \citep{Bianchi2011}, listed in the ancillary columns of the \cite{Shen2011} catalogue. 
Eightyseven percent of our sample has a measurement in at least one of the two bands and the $77\%$ has both the bands available.

The sample was originally selected from the $5^{\mathrm{th}}$ Quasar Catalogue SDSS DR7 \citep{Schneider2010}, therefore our sources have all the five optical SDSS magnitudes (\emph{ugriz}, from $3000$ to $9000$\AA) available. We used the SDSS \emph{psfMag} magnitudes listed in the catalogue of \cite{Shen2011}, i.e. a magnitude obtained fitting a PSF model to the source, optimal for point sources.

In the near-infrared range, we collected data from the Two Micron Sky Survey \citep[2MASS,][]{Skrutskie2006} in the J ($1.24 \, \mu$m), H ($1.66 \, \mu$m), and K ($2.16 \, \mu$m) bands. We matched the sources in our sample to the All Sky Point Source Catalogue and the Reject Table\footnote{The 2MASS Survey Point Reject Table (SPRT) contains 843,988,897 point sources from the Survey Working Databases that were not selected for inclusion in the in the All-Sky Release Catalog.}, requiring a maximum separation between the sources of $2 \arcsec$. When a source was not detected in the survey, we used the IDL routine \emph{aper} to infer an upper limit for the magnitudes on the 2MASS images. Specifically, for each band, we downloaded the 2MASS image, we evaluated the magnitude in a region of radius $4 \arcsec$ centered at the expected position of the source and we estimated the sky background in an annular region with inner radius $14 \arcsec$ and outer radius $20 \arcsec$. We repeated the evaluation of the sky background randomly selecting $1000$ circular regions at a distance smaller than $100 \arcsec$ from the nominal position of the source. We considered a detection if the flux of the source was larger than the $95^{\mathrm{th}}$ percentile of the distribution of $1000$ values for the sky background, otherwise we took the $95^{\mathrm{th}}$ percentile as a $2\sigma$ upper limit.
We then applied - to both detections and upper limits - the aperture correction corresponding to the image on which the aperture photometry was performed, as listed in the Atlas Image Information Table\footnote{\url{https://old.ipac.caltech.edu/2mass/releases/allsky/doc/sec2_6.html\#image_info}}.

We finally considered the four IR photometric bands of \emph{Wide-field Infrared Survey Explorer} \citep[\emph{WISE},][]{Wright2010}: W1 ($3.4 \, \mu$m), W2 ($4.6 \, \mu$m), W3($12 \, \mu$m) and W4($22 \, \mu$m).
We matched our sample to the \emph{WISE} All-Sky Source Catalogue within a $2 \arcsec$ matching radius and, where a match was not found, we searched the \emph{WISE} All-Sky Reject Table\footnote{The \emph{WISE} All-Sky Reject Table contains 284 million detections that were not selected for inclusion in the Source Catalog (either low signal-to-noise ratio or spurious detections of image artifacts)}. Ninety-nine percent of the sample has three out of the four photometric bands available and only the $13\%$ of the sample has the W4 band missing. 
We used the \emph{WISE} profile-fitting magnitude, the optimal choice for point-like sources.
The \emph{WISE} All-Sky Source Catalogue already provides upper limits when the sources are not detected. In the whole sample we only have a few tens of sources for which a measure is not available at all.

All the photometric data are corrected for Galactic reddening, using the $E(B-V)$ values as estimated from the \cite{Schlegel1998} map, using a $R_{V}=3.1$ and the extinction law by \cite{Fitzpatrick1999}.

\section{Data analysis} \label{sec:data_analysis}

\renewcommand{\arraystretch}{1.2}
\begin{table*}
	\centering
	\caption{Bins of EW[\oiii]. We divided the sample in equally spaced bins in log(EW[\oiii]), the same as in Paper I. * The first two bins have been merged in a single one. For each bin, we list mean values and uncertanties assuming a gaussian distribution and median and 16$^{\mathrm{th}}$ and 84$^{\mathrm{th}}$ percentiles (1$\sigma$ interval) for log(M$_{\mathrm{BH}}$), log(L$_{5100}$), log(L$_{\mathrm{bol}}$) and for the Eddington ratio, based on the data from the catalogue of \protect\cite{Shen2011}.}
	\label{tab:1}

	\begin{tabular}{C{1.1cm} C{0.4cm} C{0.4cm} C{1.6cm} C{1.2cm} C{1.7cm} C{1.2cm} C{1.7cm} C{1.2cm} C{1.7cm} C{0.8cm}} 
		\hline
		$\Delta$EW & EW & n & \multicolumn{2}{c}{$\mathrm{log(M}_{\mathrm{BH}}/\mathrm{M}_{\odot})$} & \multicolumn{2}{c}{$\mathrm{log(L}_{5100}/\mathrm{erg s}^{-1})$} &  \multicolumn{2}{c}{$\mathrm{log(L}_{\mathrm{bol}}/\mathrm{erg s}^{-1})$} & \multicolumn{2}{c}{$\mathrm{R}_{\mathrm{Edd}}$}  \\ \relax
		[\AA]      &  [\AA] &    &  mean   &  median    &  mean   &  median   &  mean    &  median    &  mean  & median  \\
		\hline
$1-6 ^{\,*}$& $3.0$ & $1095$&$8.392 \pm 0.017$ &$8.4^{+0.5}_{-0.5}$& $44.606 \pm 0.012$ &$44.5^{+0.4}_{-0.3}$&$45.571\pm0.012$ &$45.5^{+0.4}_{-0.3}$& $-0.921\pm0.017$&$-0.8^{+0.5}_{-0.7}$   \\
$6-12$        & $9.0$ & $3443$&$8.433 \pm 0.008$  &$8.4^{+0.5}_{-0.5}$& $44.601 \pm 0.005$ &$44.6^{+0.4}_{-0.3}$& $45.566\pm0.006$&$45.5^{+0.4}_{-0.3}$& $-0.967\pm0.008$&$-0.9^{+0.4}_{-0.5}$  \\
$12-25$      & $18.5$ & $4389$&$8.474 \pm 0.007$ &$8.5^{+0.5}_{-0.5}$& $44.587\pm 0.005$&$44.6^{+0.4}_{-0.3}$& $45.553\pm0.005$&$45.5^{+0.4}_{-0.3}$& $-1.021\pm0.007$&$-1.0^{+0.4}_{-0.5}$\\
$25-50$      & $38.5$ & $2375$&  $8.526 \pm 0.010$ &$8.5^{+0.5}_{-0.5}$&  $44.561\pm0.007$&$44.5^{+0.4}_{-0.3}$& $45.526\pm0.007$&$45.5^{+0.4}_{-0.3}$&$-1.101\pm0.009$&$-1.1^{+0.4}_{-0.5}$\\
$50-100$    & $75.0$ & $810$&  $8.520 \pm 0.016$   &$8.5^{+0.5}_{-0.5}$&  $44.519\pm0.012$ &$44.5^{+0.4}_{-0.3}$& $45.483\pm0.012$&$45.5^{+0.4}_{-0.3}$&$-1.137\pm0.015$&$-1.1^{+0.4}_{-0.4}$\\
$100-250$  & $175.0$ & $190$&  $8.466 \pm 0.039$    &$8.4^{+0.6}_{-0.5}$&  $44.470\pm0.021$&$44.4^{+0.4}_{-0.3}$ &$45.437\pm0.021$ &$45.4^{+0.4}_{-0.3}$&$-1.129\pm0.038$&$-1.1^{+0.5}_{-0.6}$\\		
		\hline
	\end{tabular}

\end{table*}
\renewcommand{\arraystretch}{1.0}

\begin{figure*}
	\includegraphics[width=2.0\columnwidth]{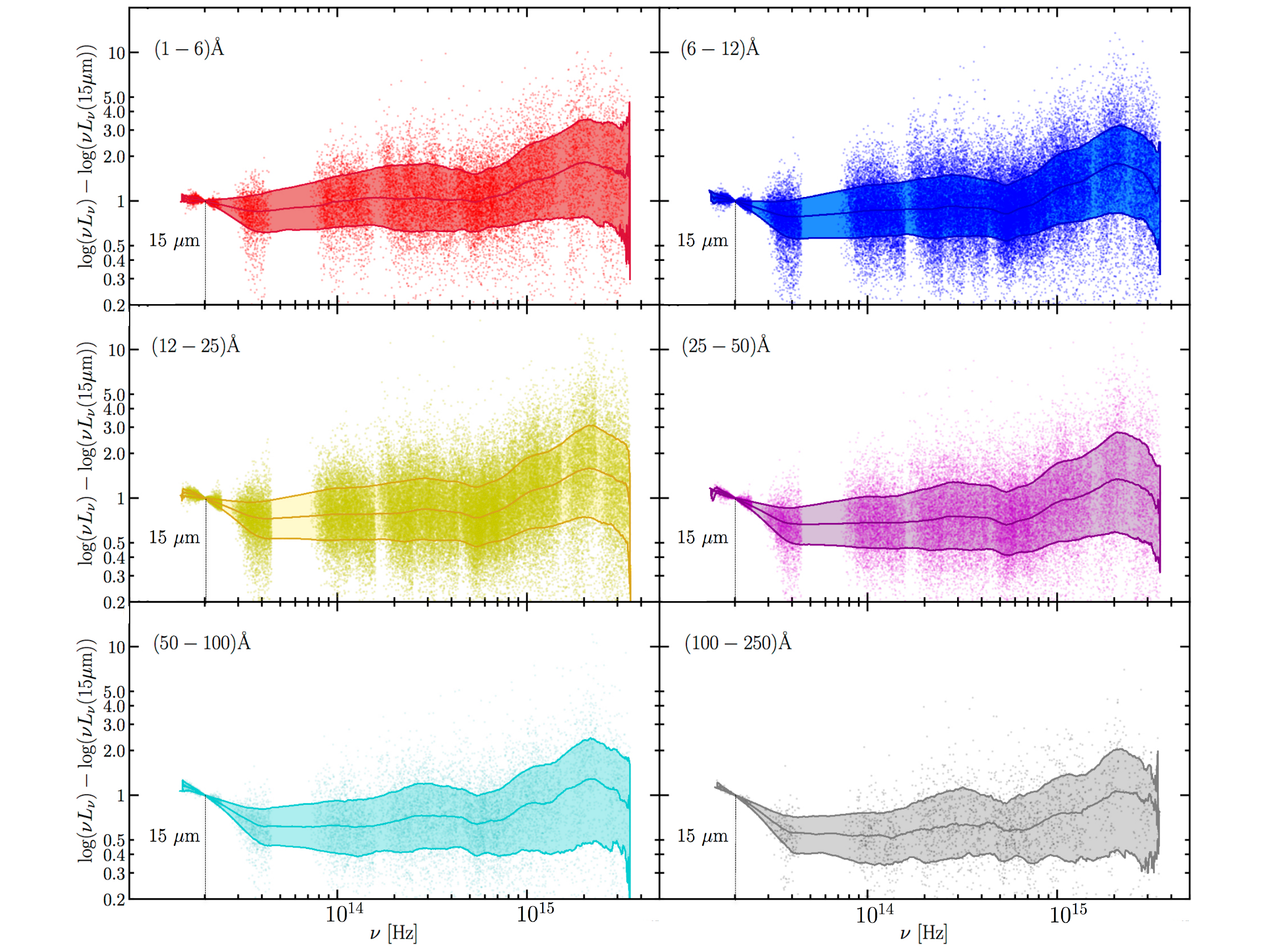}
\caption{Rest-framed photometric data points for each source in the sample (points) in six EW[\oiii] intervals, normalised to the flux corresponding to the reference wavelength $\lambda=15 \mu$m. The solid lines are the 84$^{th}$, 50$^{th}$ (median) and 16$^{th}$ percentiles - in each spectral channel - of the distribution of the SED obtained by interpolating the photometric data points for each source.}
    \label{fig:1}
\end{figure*}

To verify whether quasars SED have a trend with the EW[\oiii], we stacked the photometric data similarly to what we did with the optical spectra in Paper~I: 1) we divided the sample in six (equally spaced in log) EW[\oiii] bins (Tab. \ref{tab:1}); 2) within each bin, we rest-framed the data points accordingly to the improved measurements for the redshifts listed in the SDSS-DR7 catalogue \citep{HewettWild2010}; 3) we performed a linear interpolation of the photometric data points for each source; 4) we normalised the interpolated SED to the value corresponding to a reference wavelegth ($\lambda_{\mathrm{ref}}=15 \, \mu$m\footnote{$\lambda_{\mathrm{ref}}$ is a compromise between a representative wavelength for the torus emission and the coverage of the four \emph{WISE} photometric points, given the distribution in redshift of the sample.}) in the mid-IR, representative of the spectral range in which the torus emits. Differences in the IR emission can be ascribable to the intrinsic properties of the individual sources, both of the torus and of the primary source. Through this normalisation, we can make a comparison between the sources that is unbiased by these intrinsic differences. \footnote{Since the maximum wavelength available is $W4=22 \mu$m, for redshift higher than $z=0.47$ the reference wavelength $\lambda = 15 \mu$m falls outside of the spectral coverage. In this case, the normalisation value has been retrieved using a linear extrapolation that follows the derivative of the last two photometric points available.}; 5) we stacked all the SED in a given EW[\oiii] bin to obtain a ``representative'' SED for a given inclination angle range. In order to do that, all the SED were interpolated on a common wavelength grid and, for each spectral channel, the median value was considered.
Fig. \ref{fig:1} shows the rest-framed, photometric data (points) for all the sources in the sample, divided in the six EW[\oiii] bins. The solid lines are, from top to bottom, the 84$^{th}$, 50$^{th}$ (median) and 16$^{th}$ percentiles - in each spectral channel - of the distribution of the SED obtained by interpolating the photometric data points for each source and then normalising to the flux corresponding to the reference wavelength of $15 \mu$m.

\section{Results} \label{sec:results}

\begin{figure*}
	\includegraphics[width=2.0\columnwidth]{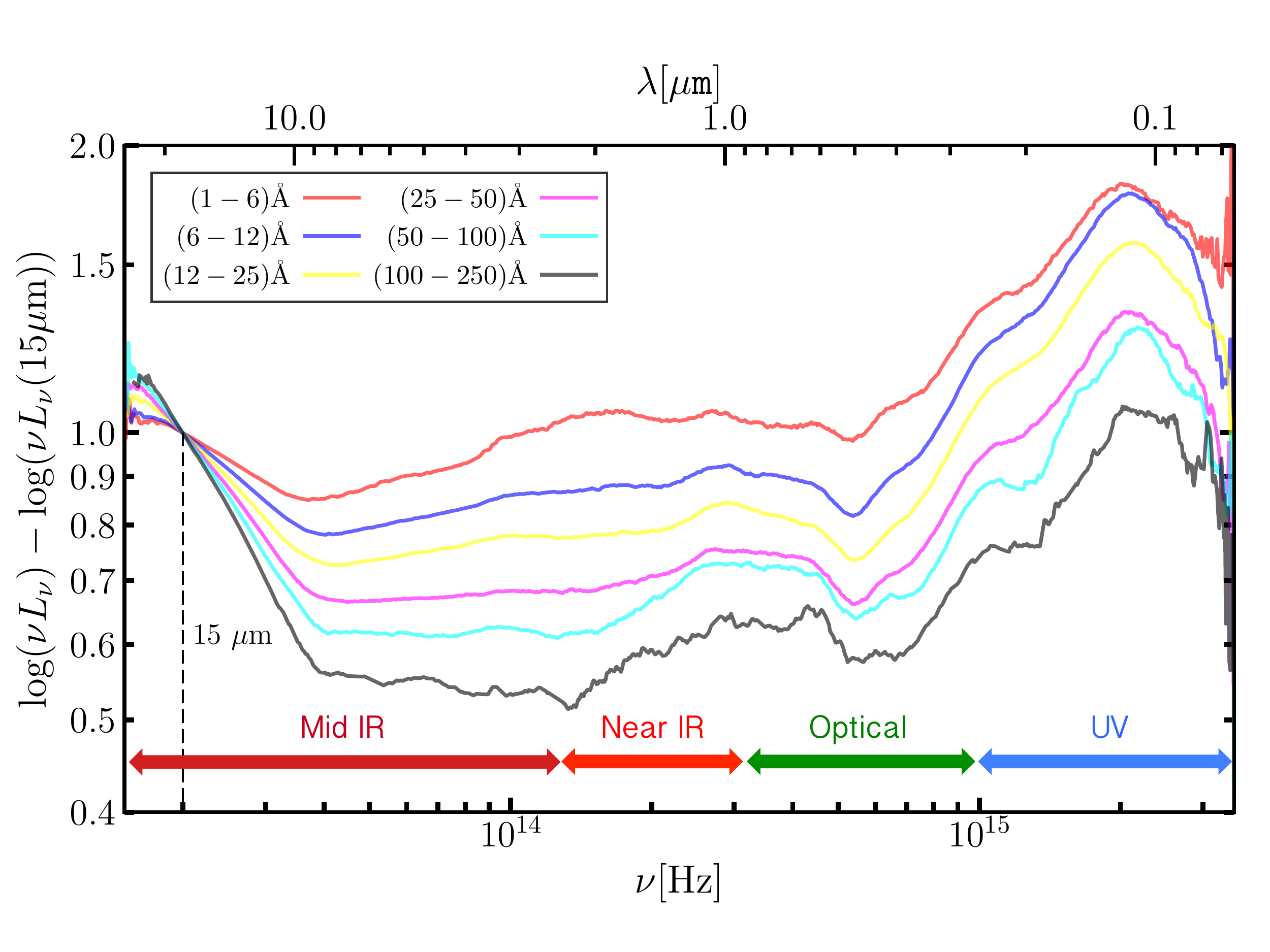}
    \caption{Comparison of the stacked SED obtained from the photometric data points for the six EW[\oiii] bins. Moving to high EW[\oiii], we see the IR part of the SED steepening, as expected in a clumpy torus scenario.}
    \label{fig:2}
\end{figure*}

\begin{figure}
	\includegraphics[width=1.0\columnwidth]{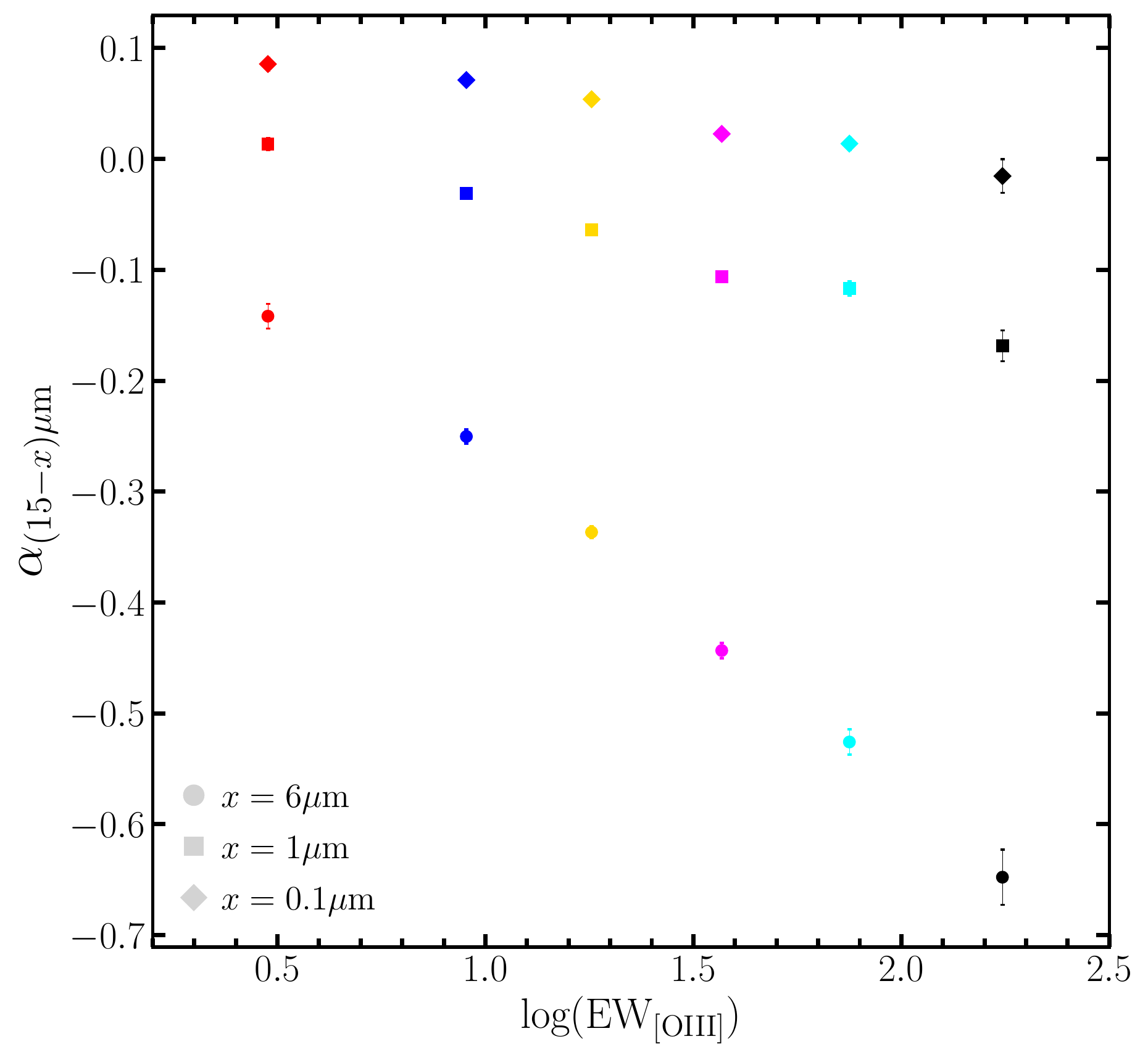}
    \caption{Energy slopes, as defined in Eq. \ref{eq:1}, for the six stacked rapresentative SED. Circles, squares and diamonds indicate the slope (15--6)$\mu$m, (15--1)$\mu$m and (15--0.1)$\mu$m respectively.}
    \label{fig:3}
\end{figure}

Fig. \ref{fig:2} compares the stacked SED ($\nu L_{\nu}$) for the six EW[\oiii] bins. The accretion disc emission (the so-called Big Blue Bump) is recognisable in the UV/optical part of the SED, the contribution of the host galaxy in the optical/near-IR range, and in the mid-IR we can see the onset of the IR Bump, related to the emission of the torus. The IR bump can be produced also by synchrotron emission coming from the most extreme among the jetted objects (blazar-like). Since our sample is composed by both radio-loud\footnote{Radio-loud objects are defined as the ones with $R=L_{6\mathrm{cm}}/L_{2500\mathrm{\AA}} > 10$. $R$ is listed in the catalogue by \cite{Shen2011}.} (994/12302 sources, 8\% of the sample) and radio-quiet quasars (11308/12302 sources, 92\% of the sample), we checked that the presence of this feature in the representative SED were not due to the presence of blazars in the sample, performing the same analysis on the sub-sample composed only by radio-quiet objects and finding exactly the same result.
The behaviour of the SED as a function of EW[\oiii] gives us two main results.
First, when we move to high EW[\oiii] values, the mid-IR part of the stacked SED steepens, with a progressive decrease of the emission at the shorter wavelengths.
Each EW bin is populated similarly in terms of luminosity, accretion rate, SMBH mass and there is no statistically significant difference among those parameter values in the different bins (see Tab. \ref{tab:1}), therefore the changes in SED shape cannot be attributed to these characteristics of the AGN components.
Second, SED corresponding to low EW[\oiii] have a more intense UV bump (of a factor of $\sim$2) than those corresponding to high EW[\oiii]. The UV emission comes from a geometrically-thin accretion disk; moving to high inclination angle (high EW[\oiii]), the intrinsic emission observed if the disc is seen face-on should decrease by a factor proportional to the $\cos \theta$.
Both these results confirm the EW[\oiii] as an inclination indicator for quasars (both radio-quiet and radio-loud).
To measure these effects, we calculated the slopes of the SED between the reference wavelength, 15$\mu$m, and 6, 1 and 0.1$\mu$m, the ratio between the energies in the IR, optical and UV bands, respectively, in the different stacks, defined as
\begin{equation} \label{eq:1}
\alpha_{(15-x)\mu\mathrm{m}} = \frac{\mathrm{log}(\nu L_{\nu_{15\mu\mathrm{m}}}) - \mathrm{log}(\nu L_{\nu_{x\mu\mathrm{m}}}) }{\mathrm{log}(\nu_{15\mu\mathrm{m}}) - \mathrm{log}(\nu_{x\mu\mathrm{m}})}\,.
\end{equation}
The significance of the difference in the slopes of the first and last stacked SED are of 25, 12 and 8 $\sigma$ - assuming a gaussian distribution - for the (15-6), (15-1) and (15-0.1)$\mu$m respectively. These results are shown in Tab. \ref{tab:2} and Fig. \ref{fig:3}.

\renewcommand{\arraystretch}{1.3}
\begin{table}
	\centering
	\caption{Energy slopes, as defined in Eq. \ref{eq:1}, for the six stacked rapresentative SED.}
	\label{tab:2}

	\begin{tabular}{C{1.05cm} C{0.4cm}  C{1.6cm} C{1.6cm} C{1.6cm}} 
		\hline
$\Delta$EW&  EW  &$\alpha_{(15-6)\mu\mathrm{m}}$&$\alpha_{(15-1)\mu\mathrm{m}}$&$\alpha_{(15-0.1)\mu\mathrm{m}}$\\ \relax
[\AA]         &[\AA]&                  &                   &                    \\
		\hline
$1-6 ^{\,*}$& $3.0$  &  $-0.142_{-0.011}^{+0.010}$ &  $+0.013_{-0.006}^{+0.006} $ & $+0.086_{-0.006}^{+0.004} $  \\
$6-12$        & $9.0$  & $-0.250_{-0.006}^{+0.005}$ &  $-0.031_{-0.003}^{+0.003} $ & $+0.071_{-0.002}^{+0.002} $  \\
$12-25$      & $18.5$ & $-0.337_{-0.005}^{+0.005}$ &  $-0.063_{-0.003}^{+0.003} $ & $+0.053_{-0.002}^{+0.002} $   \\
$25-50$      & $38.5$ & $-0.443_{-0.007}^{+0.007}$ &  $-0.106_{-0.004}^{+0.004} $ & $ +0.023_{-0.003}^{+0.003}$   \\
$50-100$    & $75.0$ & $-0.526_{-0.011}^{+0.012}$ &  $-0.117_{-0.007}^{+0.006} $ & $+ 0.014_{-0.006}^{+0.004} $   \\
$100-250$  & $175$  & $-0.65_{-0.02}^{+0.02}$     &  $-0.169_{-0.014}^{+0.015} $ & $-0.015_{-0.015}^{+0.007} $   \\		
		\hline
	\end{tabular}

\end{table}
\renewcommand{\arraystretch}{1.0}

\section{Discussion}
\label{sec:discussion}
We have demonstrated that the EW[\oiii] parameter computed from low/moderate resolution spectra (i.e. 1500 at 3800\AA) can be used as a proxy for the inclination of the accretion disc in quasars. 
Of the three possible explanations for the detection of broad line emission in quasars at high inclination angles (see Section~\ref{sec:introduction}), only one is plausible.
This is due to two factors: 1. we observe a mid-IR emission and therefore the torus can not be missing, 2. the decrease in the IR emission at shorter wavelengths as a function of EW[\oiii] is progressive - as it is with the increase in the broad lines width in paper~I; therefore a misaligned torus with respect to the inner components, i.e. BLR and accretion disc, is excluded. The only possibility is that we can intercept the emissions coming from the BLR \emph{despite} the presence of the torus, i.e. the torus \emph{must} be clumpy. 
In the following, we discuss how the behaviour of the SED as a function of EW[\oiii] is exactly what is expected in a clumpy torus scenario.

\begin{figure}
\begin{center}
	\includegraphics[width=1.0\columnwidth]{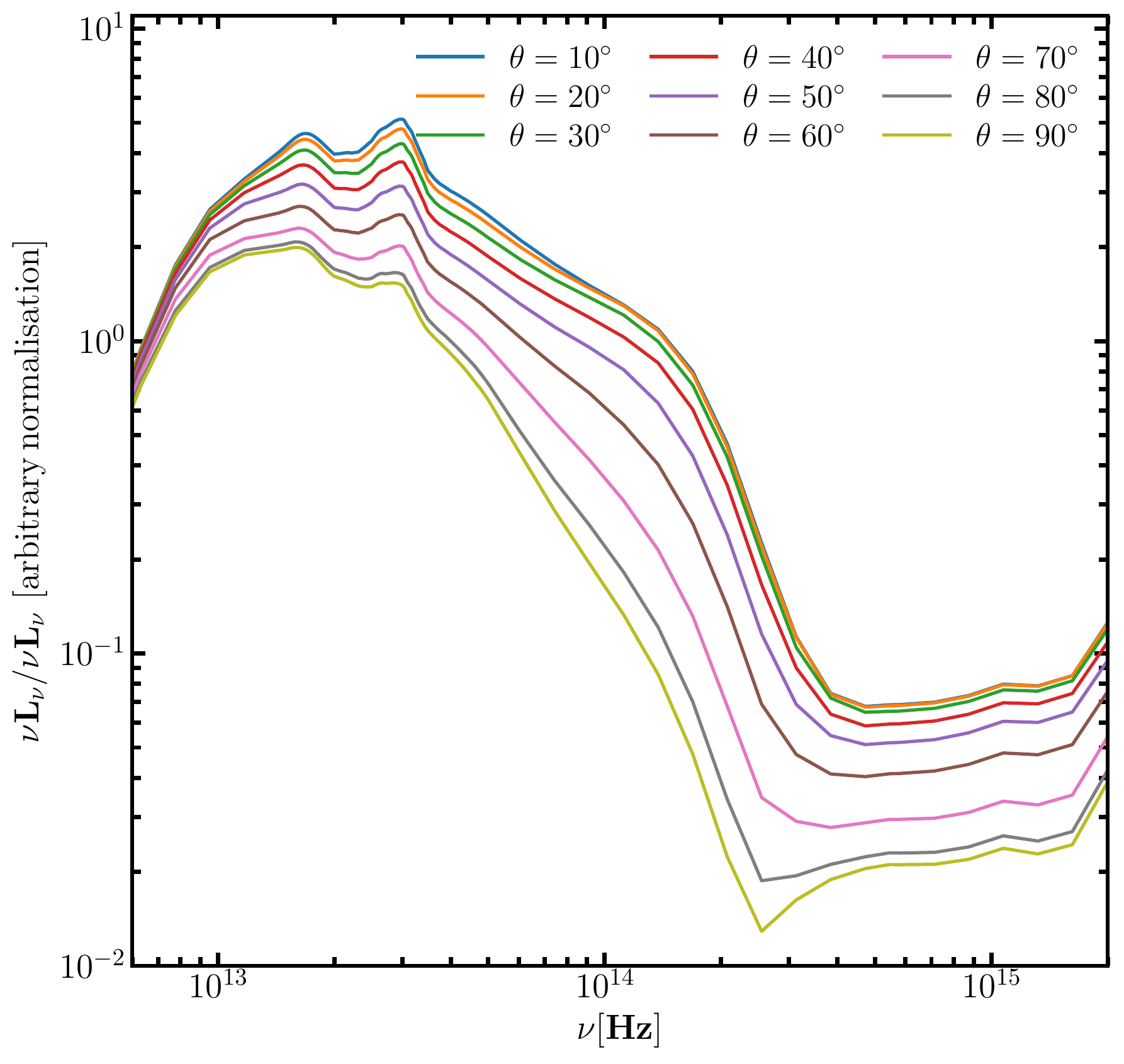}
	\caption{CLUMPY models as a function of the inclination angle of the torus with respect to the line of sight. The CLUMPY library contains more than $10^{6}$ theoretical SED, obtained varying the set of parameters describing the physics and geometry of the torus.}
    \label{fig:4}
   \end{center}
\end{figure}

The most recent models for the obscuring structure suggest that it is to be composed by clouds of molecular gas and dust \citep{Nenkova2008a, Nenkova2008b, Nikutta2009}, in some cases permeated by a diffuse medium ``filling'' the empty regions between the clouds \citep{HoenigKishimoto2010, Stalevsky2012, Siebenmorgen2015, HoenigKishimoto2017}.
As already mentioned, the clumpiness of the torus is a condition we need in order to explain the results of Paper I: we are able to intercept the emissions coming from the BLR even in sources close to the edge-on position, something that could not be explained with a smooth, optically thick torus. We want to examine the emission in the IR band as a function of EW[\oiii] to see if it is in agreement with what is expected by the clumpy models.
We considered the models by \cite{Nenkova2008a, Nenkova2008b, Nikutta2009}, which depict the torus as a toroidal distribution of clouds, whose emission, result of the re-processing of the emission of the accretion disc, can be described with a set of parameters, among which the inclination with respect to the line of sight of the observer.
The IR emission we can observe from a face-on torus is overall higher than the one coming from a torus in an edge-on position, due to the fact that in the first case we are looking at a larger portion of the emitting structure.
However, as the intensity of the IR emission can not be associated with the inclination of the accretion disc (it can be due to intrinsic differences in the properties of the individual source, such as the dimension and the covering factor of the obscuring structure, the ratio between its inner and outer radius, the properties of the emission coming from the primary source etc.), to define a reference wavelength in normalising the SED in the various bins, we have chosen a near-IR wavelegnth (i.e. 15$\mu$m). One of the most important assumption in our work is that the distribution of the tori in each bin is similar, i.e the physical properties of the torus, the dust distribution, the number of clouds, etc. are similar on average amongts the EW[\oiii] bins.
These models predict a drop in the IR emission at the shorter wavelengths ($\lambda \sim 1 \mu$m) with respect to the longer ones ($\lambda > 5 \mu$m) moving from low to high inclination angles. This result is due to the combination of two different effects: 1. going towards edge-on positions we are intercepting a higher number of clouds and 2. the absorption caused by the clouds is more efficient at the shorter with respect to the longer wavelengths. Regardless of the intrinsic distribution of the properties of the individual sources, moving from face-on to edge-on positions, the shape of the IR emission progressively steepens, with a stronger decrease in the ``blueward'' with respect to the one in the ``redward'' part.
This effect can be seen in Fig. \ref{fig:4}, where we show the shape of the torus SED as a function of the line of sight of the observer as obtained using the theoretical SED in the CLUMPY library\footnote{https://www.clumpy.org} \citep{Nenkova2008a, Nenkova2008b, Nikutta2009}. This comprehensive database contains more than 1 million of models for type~1 and type~2 AGN SED obtained varying a set of six parameters\footnote{$N_{0}$, the average number of clouds that intercept the line of sight between the primary source and the observer, $R_{o}/R_{d}$, the ratio between outer and inner radius of the toroidal structure, the second one defined by the dust sublimation radius, $q$, the power of the exponential law describing the radial density of the distribution of clouds, $\sigma$, the width parameter of the gaussian distribution of clouds with respect to the equatorial plane, $\tau_{v}$, the optical depth of each single cloud, $i$, the inclination angle} describing the physical and geometrical properties of the torus component, among which the inclination of the nucleus with respect to the line of sight.
CLUMPY also provides derived properties, such as the probability of having an unobscured AGN based on the aperture angle of the torus, as measured from the equatorial plane of the nucleus, and the inclination angle. To create this set of templates we did not exclude type~2 AGN, as defined by this parameter, since in our interpretation we can have a type~1 AGN even with a high inclination, i.e. with $i>45^{\circ}$, as long as the line of sight is free by intervening clouds.
The aim of this study is to examine the variation of the torus SED as a function of the inclination angle $i$.
We then create these templates considering, in each spectral channel, the median value of the whole set of templates with a given inclination angle (see Fig.~\ref{fig:4}), so to allow a comparison of the SED for our data to the theoretical models.
Moving from a small to a high inclination angle, we can see the decrease in the IR emission in the shorter wavelengths with respect to the longer ones.

This is exactly what we see in our data, once we compare the representative SED for the EW[\oiii] bins (Fig. \ref{fig:2}).
Moreover, the gradual steepening of the mid-IR SED with increasing inclination angle, as opposed to the bimodal transition in the SED shape predicted by a sharp-edge geometry of the distribution of the clouds \citep{Nenkova2008b}, suggests that a Gaussian angular distribution of the clouds is more representative of the obscuring structure.

We tried to fit these models to the average SEDs, but the ones corresponding to low inclination angles are still too steep to provide a good representation of the quasar stacks.
The torus models are missing an extra component that takes into account the emission produced by the hottest dust in the torus, the inner edge of the dust cloud distribution - and discussed in the most recent works on the IR SED in quasars and AGN \citep{Mor2009, Deo2011, MorNetzer2012, GarciaGonzalez2017, HoenigKishimoto2017, Duras2017, Shangguan2018}. Nonetheless, we clearly see the decrease in the shortest wavelengths emission of the IR SED moving from low to high EW[\oiii] (see Tab. \ref{tab:2}). Given that in each EW[\oiii] the sources are similarly distributed in terms of BH mass, luminosity and Eddington ratio (Tab. \ref{tab:1}), we can not find any other physical explanation to this trend other than the effect produced by the orientation of the torus.

As already stressed in paper~I, we do not claim the orientation to be the only driver in the variance of quasars spectra in general, or in the IR SED for the specific case of the torus. Other physical drivers may contribuite to make the physical properties and the geometry of the system - and therefore the SED - different. In the case of the torus, both primary source properties - such as the accretion rate \citep{Ricci2017, Ezhikode2017} or the luminosity \citep{Maiolino2007, Treister2008, Lusso2013}, and composition of the torus clouds \citep{Audibert2017, RamosAlmeida2011, AlonsoHerrero2011} can change the shape of the IR emission.
We do note again, however, that, as for this study has been designed, the effects produced by drivers other than orientation are ``blended'' within the EW[\oiii] bins. The effect produced by the orientation is the only one recognisable in the stacking we performed.

We are undertaking a complete SED fitting to our data using AGNfitter \citep{CalistroRivera2016}, a Bayesian MCMC code that takes into account four emitting components for reproducing the total emission of AGN: the accretion disc, responsible for the Big Blue Bump in the optical/UV, the hot dust torus, the stars in the host galaxy and the cold dust in star forming regions. As already mentioned, a preliminary analysis on our sample has revealed that, to properly treat the effect produced by the orientation on quasars SED, we need to include an extra black body component.
This will be the subject of a forthcoming paper.

\section{Conclusions}

In this work we constructed the broad-band SED of $\sim 12,300$ SDSS DR7 quasars and analysed the overall shape as a function of the inclination angle of the accretion disc with respect to the line of sight. The EW[\oiii] orientation indicator allows us to split the sample in bins of different inclination angles and examine the presence of orientation effects on the SED of quasars.
In previous works on the EW[\oiii] orientation indicator, we found that broad emission lines are detectable even in sources close to an edge-on position, in agreement with the most recent models depicting the torus as a clumpy rather than a smooth structure. In this sense, type~1 AGN are sources for which the line of sight is free from intervening clouds, regardless of their inclination. \\
We focus on the IR bands of the SED - related to the emission from the obscuring torus - looking for information on the properties of this structure in light of its inclination with respect to the observer. Our findings are the following.
\begin{itemize}
\item[1.] We do see a clear trend of the representative SED we created dividing the sample in EW[\oiii] bins. Our sample is composed by objects with every kind of luminosity, accretion rate, BH mass and other properties of AGN components, and since it has been divided only in terms of EW[\oiii], a quantity that is defined to be a function of the inclination of the accretion disc with respect to the line of sight, we interpret the behaviour of the representative SED as a confirmation of the reliability of this quantity as an orientation indicator.
\item[2.] We do detect an emission in the mid-IR wavelenght range, therefore the torus can not be missing. To check that the IR bump is not produced by blazar-like objects, we performed the stacking analysis on the 8\% and 92\% of quasars that show strong radio emission and that are radio-quiet respectively, and found the same results.
\item[3.] The IR SED {\it \bf gradually} steepens moving from low EW[\oiii] (face-on) to high EW[\oiii] (edge-on) objects (Tab. \ref{tab:2} and Fig. \ref{fig:3}). EW[\oiii] is a strong function of the inclination of the geometrically-thin accretion disc. Given that the energy slopes decrease so neatly with EW[\oiii], it means that the obscuring structure on a statistical sense has then to be co-axial with accretion disc and BLR. Our results also imply that the torus has a clumpy structure. From the results of Paper~I we know that we can intercept the broad emissions even in sources in a position close to the edge-on; the torus has therefore to be composed by clumps rather than by a smooth continuum of gas, and our ability of intercepting the broad lines is related to the probability of having a line of sight free by intervening clouds.
\item[4.] The decrease of the IR emission at the shorter with respect to the longer wavelengths is in agreement with what expected from the theoretical models for clumpy tori.
\end{itemize}
We conclude that the only possible explanation for type~1 objects in position close to the edge-on, i.e. for the detection of broad emission lines, is a torus with a clumpy structure.
The use of EW[\oiii] as an orientation indicator for quasars allows us to verify the presence of orientation effects on the IR SED.
The orientation plays a major role in determining the shape of the SED in quasars, although it is not the only driver. Other physical drivers (e.g. SMBH mass, luminosity, Eddington ratio) may compete in shaping quasars emission. This work is however designed to isolate the effects produced by the orientation; in each inclination range we are averaging sources that span a wide range in terms of luminosity, accretion rate and composition.

\section*{Acknowledgements}

S.B. and G.R. acknowledge support from the grant ASI-INAF N. 2017-14-H.0.
S.B. is supported by NASA through the Chandra award no. AR7-18013X issued by the ChandraX-ray Observatory Center, operated by the Smithsonian Astrophysical Observatory for and on behalf of NASA under contract NAS8-03060. S.B. was also partially supported by grant HST-AR-13240.009.  
E.L. is supported by an EU COFUND/Durham Junior Re- search Fellowship under grant agreement no. 609412.
Funding for the SDSS and SDSS-II has been provided by the Alfred P. Sloan Foundation, the Participating Institutions, the National Science Foundation, the U.S. Department of Energy, the National Aeronautics and Space Administration, the Japanese Monbukagakusho, the Max Planck Society, and the Higher Education Funding Council for England. The SDSS Web Site is http://www.sdss.org/.
The SDSS is managed by the Astrophysical Research Consortium for the Participating Institutions. The Participating Institutions are the American Museum of Natural History, Astrophysical Institute Potsdam, University of Basel, University of Cambridge, Case Western Reserve University, University of Chicago, Drexel University, Fermilab, the Institute for Advanced Study, the Japan Participation Group, Johns Hopkins University, the Joint Institute for Nuclear Astrophysics, the Kavli Institute for Particle Astrophysics and Cosmology, the Korean Scientist Group, the Chinese Academy of Sciences (LAMOST), Los Alamos National Laboratory, the Max-Planck-Institute for Astronomy (MPIA), the Max-Planck-Institute for Astrophysics (MPA), New Mexico State University, Ohio State University, University of Pittsburgh, University of Portsmouth, Princeton University, the United States Naval Observatory, and the University of Washington.
This publication makes use of data products from the Two Micron All Sky Survey, which is a joint project of the University of Massachusetts and the Infrared Processing and Analysis Center/California Institute of Technology, funded by the National Aeronautics and Space Administration and the National Science Foundation.
This publication makes use of data products from the Wide-field Infrared Survey Explorer, which is a joint project of the University of California, Los Angeles, and the Jet Propulsion Laboratory/California Institute of Technology, funded by the National Aeronautics and Space Administration.
This work made use of matplotlib, a Python library for publication quality graphics \citep{Hunter2007}, and of the software for the analysis and manipulation of catalogues and tables TOPCAT \citep{Taylor2005}, available online at the link \url{http://www.starlink.ac.uk/topcat/}.




\bibliographystyle{mnras}
\bibliography{torus_OIII_ArXiv} 


%
%
%
%
%

\bsp	
\label{lastpage}
\end{document}